\documentclass[superscriptaddress,twocolumn,showpacs,preprintnumbers,amsmath,amssymb,nofootinbib,aps]{revtex4}
\usepackage{graphicx}
\usepackage{dcolumn}
\usepackage{bm}
\usepackage{url}

\begin{document}

\title{A Deeper Look at Student Learning of Quantum Mechanics: the Case of Tunneling}


\author{S. B. McKagan}
\affiliation{JILA and NIST, University of Colorado, Boulder, CO, 80309, USA}

\author{K. K. Perkins}
\affiliation{Department of Physics, University of Colorado, Boulder, CO, 80309, USA}

\author{C. E. Wieman}
\affiliation{Department of Physics, University of British Columbia, Vancouver, BC V6T 1Z1, CANADA}
\affiliation{JILA and NIST, University of Colorado, Boulder, CO, 80309, USA}
\affiliation{Department of Physics, University of Colorado, Boulder, CO, 80309, USA}

\date{\today}

\begin{abstract}
We report on a large-scale study of student learning of quantum tunneling in 4 traditional and 4 transformed modern physics courses.  In the transformed courses, which were designed to address student difficulties found in previous research, students still struggle with many of the same issues found in other courses.  However, the reasons for these difficulties are more subtle, and many new issues are brought to the surface.  By explicitly addressing how to build models of wave functions and energy and how to relate these models to real physical systems, we have opened up a floodgate of deep and difficult questions as students struggle to make sense of these models.  We conclude that the difficulties found in previous research are the tip of the iceberg, and the real issue at the heart of student difficulties in learning quantum tunneling is the struggle to build the complex models that are implicit in experts' understanding but often not explicitly addressed in instruction.
\end{abstract}

\maketitle

\section{Introduction}
Tunneling is a surprising result that has served to validate the theory of quantum mechanics by explaining many real world phenomena such as alpha decay, molecular bonding, and field emission, and has resulted in applications such as scanning tunneling microscopes.  As a case study in the counterintuitive yet applicable nature of quantum mechanics, tunneling is an important part of any introductory course in modern physics or quantum mechanics.

An examination of modern physics and quantum mechanics textbooks, course syllabi, and interviews with faculty who have taught such courses, suggest that instruction in tunneling should help students achieve the following learning goals:
\begin{enumerate}
  \setlength{\itemsep}{1pt}
  \setlength{\parskip}{0pt}
  \setlength{\parsep}{0pt}
    \item Calculate or discuss qualitatively (depending on the level of the course) the probability of tunneling for various physical situations
    \item Describe the meaning of the potential energy and wave function graphs.
    \item Visualize how these graphs would change if the physical situation were altered.  (e.g. changing barrier height and width)
    \item Relate the mathematical formalism and graphical representation of tunneling to the phenomenon of tunneling in the real world.
\end{enumerate}

Tunneling has been a favorite topic of physics education researchers specializing in quantum mechanics, who have found that many students have a great deal of trouble understanding even the most basic aspects of this topic~\cite{Ambrose1999a,Bao1999a,Morgan2004a,Wittmann2005a,Falk2004a,Domert2005a,McKagan2006a}.  In designing a transformed course in modern physics for engineering majors~\cite{McKagan2007a}, we drew on the literature of previous research to develop a curriculum aimed at addressing known student difficulties in understanding quantum tunneling~\cite{2130}.  Throughout the process of developing and refining this course, we carried out a study to answer the following research questions:
\begin{enumerate}
  \setlength{\itemsep}{1pt}
  \setlength{\parskip}{0pt}
  \setlength{\parsep}{0pt}
    \item Does our transformed curriculum help to address common student difficulties in learning tunneling?
    \item Are our students achieving the learning goals described above?
    \item What are the practices that support or hinder the achievement of these goals?
\end{enumerate}

We find that our curriculum does help students overcome common difficulties and achieve our learning goals.  While the common difficulties reported in the literature do arise in the transformed classes, they are less prevalent than in comparable traditional classes, and they often arise in more subtle ways and for different reasons than discussed in the previous literature.

Our research shows that the difficulties discussed in the literature are surface features, masking a much more serious problem: In tunneling, as in other aspects of quantum mechanics, \emph{students fail to grasp the basic models that we are using to describe the world}.  These models include wave functions as descriptions of physical objects, potential energy graphs as descriptions of the interactions of those objects with their environments, and total energy as a delocalized property of an entire wave function that is a function of position.  Hestenes has pointed out that while ``A physicist possesses a battery of abstract models with ramifications already worked out or easily generated,'' standard physics instruction often treats these models implicitly, rather than explicitly.~\cite{Hestenes1987a}  While this is true even in introductory physics, the problem is more serious in quantum mechanics, where the models are particularly abstract.  Standard instruction in quantum mechanics, including tunneling, does not provide students with enough information to make sense of these models, or even recognize that they exist.  We have achieved a degree of success in teaching quantum tunneling by making these models more explicit, and suggest further changes in this direction.

\section{The standard presentation of Quantum Tunneling}
\begin{figure}[htbp]
  \fbox{\includegraphics[width=\columnwidth]{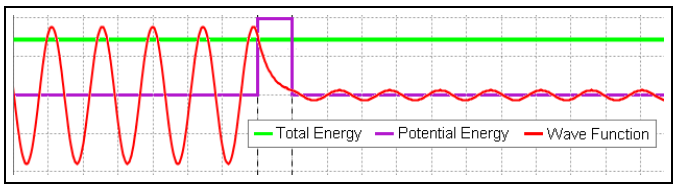}}
  \caption{\label{planewave}The standard presentation of quantum tunneling: a plane wave tunneling through a square potential barrier.  Total energy, potential energy, and the real part of the wave function are all drawn on the same graph, and the real part of the wave function is labeled as simply ``wave function.''}
\end{figure}

Most textbooks on modern physics and quantum mechanics have a discussion of quantum tunneling.  The discussion is remarkably similar throughout these books, with the main difference being that modern physics textbooks give less detail.  Tunneling is defined as a wave function passing through a potential energy barrier that is greater than its total energy.  The typical presentation includes an analysis of the plane wave solution to the Schr{\"o}dinger equation for a square potential energy barrier, as shown in Figure~\ref{planewave}.  Often the wave function, potential energy, and total energy are drawn on the same graph, a practice that research has shown to lead to student confusion~\cite{Morgan2004a,Domert2005a}.  Depending on the level of the textbook, the reflection and transmission coefficients are either derived or given.  This is typically followed by a discussion of some applications of quantum tunneling, such as alpha decay, scanning tunneling microscopes, and the inversion of Ammonia molecules.  Some textbooks also include a discussion of tunneling wave packets, occasionally showing pictures of a tunneling wave packet taken from a numerical simulation such as in Ref.~\cite{Goldberg1967a}.  Wave packets and applications are nearly always relegated to the end of the discussion of tunneling.

In examining the standard presentation of tunneling, one may ask how it aligns with the learning goals in Section I.  The standard presentation certainly gives students practice in calculating relevant quantities for the case of a plane wave and square barrier, but it does not give students the tools to extend these calculations to more realistic systems.  It also includes both a mathematical model and a discussion of physical applications of this model.  However, we argue that it does not provide sufficient links between the two.  For example, there is almost never a discussion of what physical system could produce the square barrier shown in Figure~\ref{planewave} or of how a plane wave relates to a real particle. Further, when real applications are discussed, their potential energy graphs are often not discussed, making it harder for students to relate the applications to the mathematical model.  Thus, the standard presentation does not provide students with the tools to extend the model of tunneling beyond square barriers to the more complicated potentials involved in real physical systems, either quantitatively or qualitatively.

\section{Previous Physics Education Research on Quantum Tunneling}
Many researchers have documented student difficulties in learning quantum tunneling~\cite{Ambrose1999a,Bao1999a,Morgan2004a,Wittmann2005a,Falk2004a,Domert2005a,McKagan2006a}.  These researchers, working at many institutions in the United States and Sweden, have found a fairly consistent list of student difficulties.

The most common difficulty, discussed in all these references, is the belief that energy is lost in tunneling.  The correct description of energy in quantum tunneling is that because there is no dissipation in the Schr{\"o}dinger equation, energy is conserved, as can be seen in Figure~\ref{planewave}, where the total energy is constant throughout.  The barrier itself represents the potential energy, which is zero on the left and right, and some positive constant inside the barrier.\footnote{It is linguistically awkward to speak of the potential energy ``inside the barrier'', since the potential energy \textit{is} the barrier, but it is important to be explicit, as many students do not recognize the equivalence of potential energy and barrier.}  The kinetic energy is equal to the total energy on the left and right, and is negative inside the barrier.  Ambrose~\cite{Ambrose1999a} and Bao~\cite{Bao1999a} report the student belief that \textit{kinetic} energy is lost in tunneling, although later research shows that this difficulty is not limited to kinetic energy: Morgan et al.~\cite{Morgan2004a} quote students as saying that ``energy'' is lost, without specifying which kind of energy, and in our own work, we found that most students who thought that energy is lost did not have a clear idea of \textit{which} energy is lost. When asked, they were just as likely to say potential, kinetic, or total energy, and often used two or even all three types of energy interchangeably within the same explanation.~\cite{McKagan2006a}

There are two common explanations in the literature for the belief that energy is lost in tunneling. The first explanation, attributable to the fact that most textbooks and lecturers draw the energy and the wave function on the same graph, is that students confuse the two, believing that the energy, like the wave function, decays exponentially during tunneling~\cite{Bao1999a,Morgan2004a,Domert2005a}.  The second explanation is that students think that ```work' is done on or by the particles while inside the potential barrier''~\cite{Ambrose1999a} or that energy is ``dissipated'' as in a physical, macroscopic tunnel~\cite{Morgan2004a}.  Many researchers report on student interviews showing that both these explanations are common among students.~\cite{Ambrose1999a,Bao1999a,Morgan2004a,Wittmann2005a,Falk2004a,Domert2005a,McKagan2006a}

A third possible explanation suggested by Bao is that students may be thinking of mechanical or electromagnetic waves, in which the energy of the wave \emph{is} related to the amplitude.~\cite{Bao1999a}  However, no evidence is presented to support this explanation of student thinking.  In our observations and interviews in \emph{traditional} modern physics courses, few students have sufficient understanding of mechanical or electromagnetic waves to cause problems in their interpretation of the amplitude of matter waves, and none have used such an explanation.  As discussed in the Section VIA (reason 4), we \emph{do} see some evidence of students using this explanation for energy loss in our \emph{transformed} modern physics course, in which the dependence of amplitude on energy in electromagnetic waves is heavily stressed.

\begin{figure}[htbp]
  \fbox{\includegraphics[width=\columnwidth]{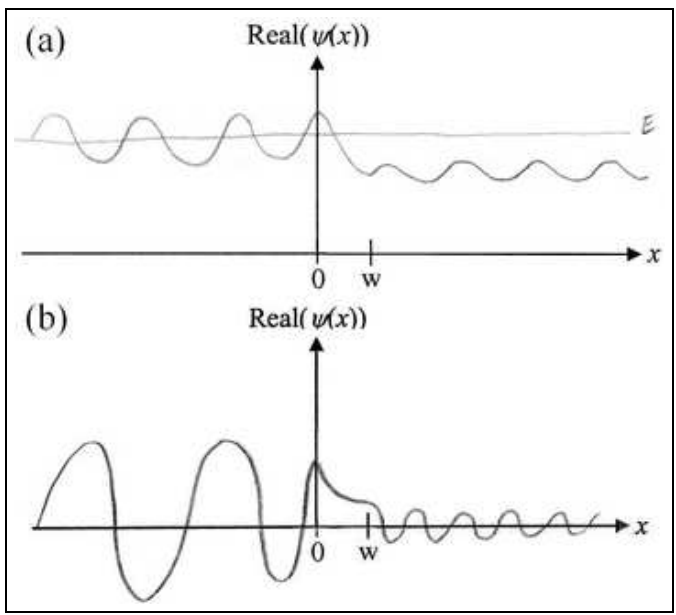}}
  \caption{\label{difficulties}Common student difficulties reported in the literature: incorrectly drawing the real part of the wave function with (a) an offset between the horizontal axes on the left and right side of the barrier and (b) a smaller wavelength on the right than on the left.  These drawings are taken from student responses to an exam question asking students to draw the real part of the wave function, as discussed in Section VI.  We have observed physics faculty making drawings similar to both (a) and (b), and a popular introductory quantum mechanics textbook contains a figure similar to (b)~\cite{Griffiths2005a}.}
\end{figure}

Other common student difficulties reported in the literature are: the belief that reflection at a barrier is due to particles having a range of energies~\cite{Ambrose1999a}; incorrectly drawing the wave function with an offset between the horizontal axes of the wave function on the left and right side of the barrier, as in Figure~\ref{difficulties}a~\cite{Morgan2004a}; incorrectly drawing the wave function with a smaller wavelength on the right than on the left, as in Figure~\ref{difficulties}b~\cite{Ambrose1999a,Morgan2004a}; and misinterpreting the meaning of the wavelength and amplitude of the wave function.

In addition to these common student difficulties, in our own previous research we found that many students do not know what the potential energy graph represents~\cite{McKagan2006a}.  Our results from student interviews are supported by many conversations with practicing physicists who report having successfully completed quantum mechanics courses as students without realizing what a potential well was until much later. We believe that this problem is due to the lack of physical context for potential energy graphs in the standard treatment discussed in Section II.  We will return to this issue later.

Brookes and Etkina~\cite{Brookes2006a,Brookes2006b} argue that physicists talk about potential using a metaphor of a physical object, as illustrated by the terms ``potential well,'' ``potential barrier,'' and ``potential step.''  Because these metaphors are implicit and their limitations are not discussed, students have a tendency to overextend them, leading to many of the student difficulties that other researchers have documented.  Brookes and Etkina's analysis overlaps with ours, in that they also point out that physics professors are not explicit in discussing the limitations of models.

\section{The Study}
In order to answer the research questions in Section I, we collected qualitative and quantitative data on student thinking about quantum tunneling in eight modern physics courses over a five-semester period.  Five of these courses were for engineering majors and three were for physics majors.  Four of the engineering majors' courses were taught using the transformed curriculum described in the next section.  The first two semesters of the transformed course were taught by the authors, and the next two by another professor in the physics education research (PER) group. The remaining courses in this study were taught using traditional methods.

The qualitative data we collected consist of observations of students in approximately 200 lectures (20 on tunneling) and 50 problem-solving sessions (5 on tunneling), student responses to essay questions on homework and exams, and student interviews.  The interviews include 44 interviews on the Quantum Mechanics Conceptual Survey (QMCS)~\cite{QMCS}, which includes questions on  tunneling~\cite{McKagan2006a}, 6 interviews on the \emph{Quantum Tunneling and Wave Packets} simulation described in the following section, and 2 interviews on tunneling with each of 6 students who participated in a case study project involving regular interviews throughout the semester.  The quantitative data consist of student responses on the QMCS, homework, and exams.

By drawing on multiple forms of data, we have been able to track similar responses among many courses, as well as looking at changes in student thinking as further transformations were introduced into the curriculum.

\section{An Improved Curriculum for Teaching Quantum Tunneling}
As part of the transformation of a modern physics course for engineering majors~\cite{McKagan2007a}, we developed a curriculum for teaching quantum tunneling.  The course design was based on PER, using interactive engagement techniques such as peer instruction and collaborative homework sessions, focusing on real-world applications, and addressing common student difficulties.  The curriculum on quantum tunneling was designed to address common student difficulties with this topic, which were known from previous research (see section III).~\cite{2130}

\subsection{Addressing student difficulties with energy loss}
Several aspects of the instruction were designed to address the belief that energy is lost in tunneling.  As discussed in section III, previous research cites two reasons that students believe energy is lost in tunneling: (1) confusion between wave function and energy, and (2) invoking dissipation.

\begin{figure}[htbp]
  \fbox{\includegraphics[width=\columnwidth]{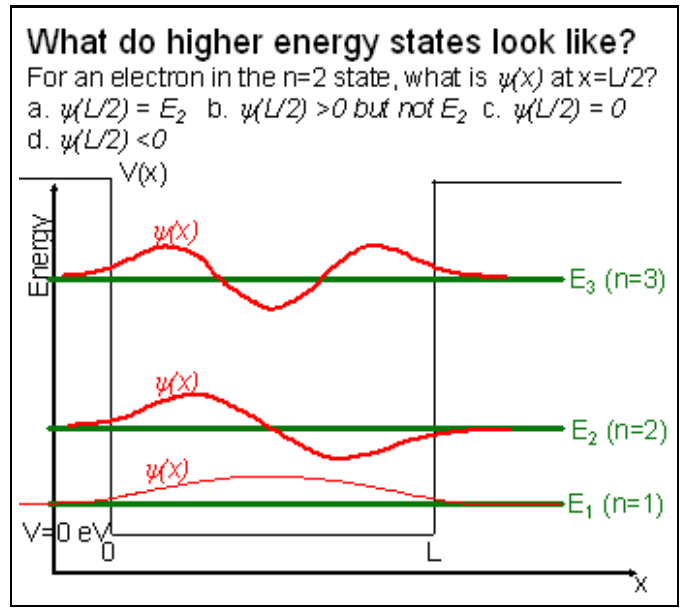}}
  \caption{\label{clicker}A concept question designed to elicit student confusion between energy and wave function.  The correct answer is C, but students who don't understand the meaning of superimposing a wave function graph on an energy graph may be inclined to answer A.  When we ask this question in class, it generates a large amount of discussion.  While most students (73-88\%, depending on the semester) eventually answer the question correctly, listening in on student discussions reveals that most don't know the answer right away, and only figure it out through vigorous debate with their neighbors.  Even after discussion, 9-19\% give answer A.}
\end{figure}

To address reason (1), we were careful to draw energy and wave function on separate graphs.  However, since the representation in Figure~\ref{planewave}, in which they are plotted on the same graph, is ubiquitous in textbooks and other literature, it is impossible to avoid students being exposed to it.  This representation has been so ingrained in us by our own education that we had to be on guard to keep from drawing graphs this way ourselves!  Therefore, we also used concept questions (multiple choice questions posed in class that students discuss in small groups and answer using a personal response system) and homework questions to elicit student confusion between energy and wave function and address it directly.  Figure~\ref{clicker} shows an example of a concept question used to address this confusion.

To address reason (2), we emphasized energy conservation and the lack of dissipation in the Schr{\"o}dinger equation.  One key feature of our curriculum was a tutorial~\cite{2130} adapted from the Quantum Tunneling Tutorial in the Activity-Based Tutorials Volume 2~\cite{Wittmann2005b}, developed by Wittmann, Steinberg, and Redish.  This tutorial was designed to address the belief that energy is lost in tunneling by asking students to work out the total, kinetic, and potential energy in each region and answer questions about energy conservation.

\subsection{Putting potential energy in context}
We also designed our curriculum to address our previous finding that students are often confused by the meaning of the potential energy function~\cite{McKagan2006a}.  We consistently gave a physical context for potential energy functions, presenting square wells and barriers as illustrations of real physical systems, rather than mere abstractions.  It is worth noting that it was a great challenge for our team of three expert physicists, including one Nobel Laureate, to think of even a single real physical system represented by a square well or a square barrier.

\begin{figure}[htbp]
  \fbox{\includegraphics[width=\columnwidth]{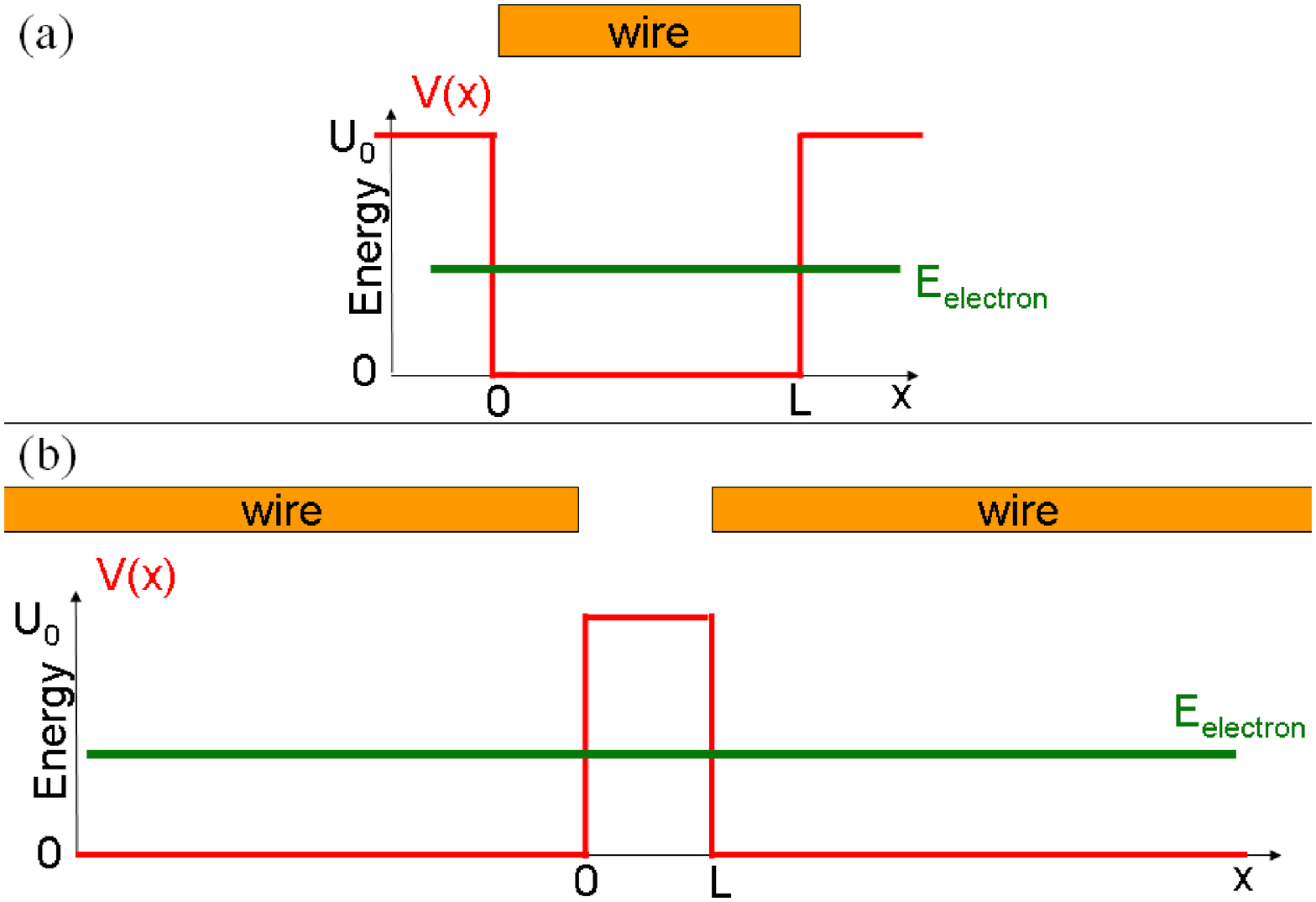}}
  \caption{\label{wires}Physical contexts for (a) a square well and (b) a square barrier.  A square well with width L and height $U_0$ represents a wire with length $L$ and work function $U_0$.  A square barrier with width L and height $U_0$ represents two long wires with work function $U_0$ separated by an air gap with length $L$.}
\end{figure}

The physical examples that we decided to use in our course are illustrated in Figure~\ref{wires}: an electron in a short wire as the context for a square well, and an electron traveling through a long wire with a thin air gap as the context for a square barrier.

We chose the physical context of an electron in a wire because it has practical applications for real circuits.  While there are a few textbooks that provide physical examples of tunneling (an electron bouncing back and forth between two capacitors with tiny holes in them for a square well~\cite{Harris1999a}, and an electron traveling through a series of metal tubes held at different voltages for a square barrier~\cite{Eisberg1985a}), these examples are so artificial that no one would ever create such a system for any reason other than to demonstrate the abstract potentials used in introductory quantum mechanics courses.  We decided against using the example of a charged bead moving along a wire held at different potentials that was used in the original version of the Activities-Based Tutorials~\cite{Wittmann2005b}, also because it seemed excessively artificial.

Our curriculum included many opportunities for students to practice building models of how potential energy graphs relate to physical systems.  For example, in interactive lectures, homework problems, and a tutorial in we asked students to build up potential energy diagrams for systems such as an electron in a wire, a scanning tunneling microscope, and a nucleus undergoing alpha decay.  We also asked students to reason through the physical meaning of the potential energy for various systems.

Further, we used the term ``potential energy,'' rather than the shorthand ``potential,'' to avoid confusion.  Although it would be preferable to use the symbol $U$, rather than the common convention $V$, for potential energy, to help students relate the potential energy in quantum mechanics to the potential energy in other areas of physics, we used $V$ in order to be consistent with the textbook we chose for the first semester.  However, we repeatedly emphasized the meaning of this symbol, and explicitly pointed out the inconsistency in notation among different areas of physics.

\subsection{The \emph{Quantum Tunneling} Simulation}
The standard presentation of quantum tunneling discussed in Section II provides an abstract and decontextualized model that is difficult to visualize or connect to reality.  The content of this presentation is artificially constrained by what can be calculated.  Students learn to calculate transmission coefficients for plane waves tunneling through square barriers, not because this is a relevant problem, but because this is the only tunneling problem that can reasonably be calculated analytically.  With modern computational techniques, however, it is no longer necessary for the curriculum to be so constrained.

\begin{figure}[htbp]
  \includegraphics[width=\columnwidth]{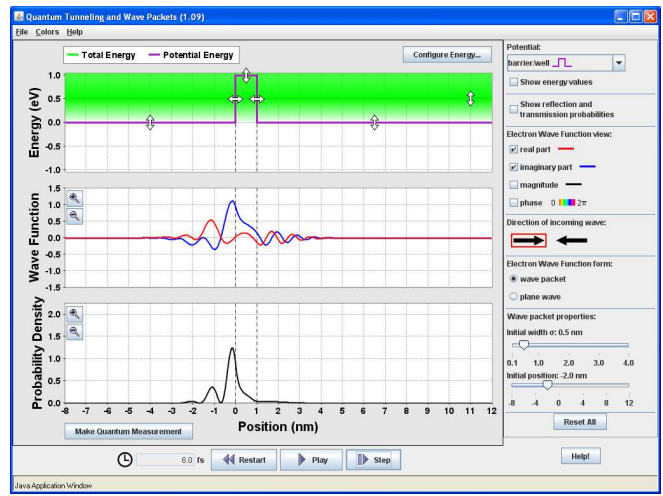}
  \caption{\label{simulation}The \emph{Quantum Tunneling and Wave Packets} simulation provides interactive visual models of tunneling of wave packets and plane waves in a variety of physical situations, and removes constraints imposed on curriculum by what problems can be easily calculated.}
\end{figure}

We designed the \emph{Quantum Tunneling and Wave Packets} simulation~\cite{QTsim} (see Figure~\ref{simulation}) to provide easily accessible interactive visual models of tunneling of wave packets and plane waves in a variety of physical situations, thus removing many constraints on curriculum.  With the simulation, we can begin our instruction with wave packets, rather than plane waves, so that students can develop a visual model of what is happening in time and space in quantum tunneling.  This simulation was developed as part of the Physics Education Technology Project (PhET)~\cite{PhET}, which provides free interactive computer simulations for teaching physics.  Like other PhET simulations, the \emph{Quantum Tunneling and Wave Packets} simulation is highly interactive, allowing students to change the potential and total energies by dragging on the graph, so that they can quickly explore a wide variety of physical situations that would be cumbersome to calculate.  The simulation also provides a wide variety of representations, allowing students to view the real part, imaginary part, magnitude, and phase of the wave function.  To address the problem of students mixing up energy and wave function, these quantities are displayed on separate graphs in the simulation.

\section{Results}

Even in our transformed courses, we saw some evidence of most of the difficulties reported in previous research on student understanding of quantum tunneling.  For example, in a final exam question in which students were asked to draw the real part of the wave function for an electron tunneling through a square barrier and explain their reasoning, $18\%$ of students drew the wave function with an offset as in Figure~\ref{difficulties}a, and $23\%$ drew a shorter wavelength on the right than on the left as in Figure~\ref{difficulties}b.  However, many of the previously reported difficulties were less prevalent or appeared in more subtle forms than we saw in traditional courses.  We also saw many issues in our transformed courses that have not been previously reported.

\subsection{Energy Loss: a new perspective}
In the transformed courses, because there was such a heavy emphasis on energy conservation, students quickly learned to say that energy is not lost in tunneling.  When we asked them directly on exams whether ``the total energy of an electron after it tunnels through a potential barrier is a) greater than, b) equal to, or c) less than its energy before tunneling,'' between $70\%$ and $93\%$ answered correctly that it is equal.  In homework, when students were asked ``Does an electron lose energy when it tunnels?'' between $95\%$ and $96\%$ answered correctly that it does not and gave clear explanations of their reasoning, invoking the conservation of energy and the lack of dissipation.

\begin{table}
\begin{tabular*}{\columnwidth}{|p{.982\columnwidth}|}
  \hline
  Reasons students may think energy is lost in tunneling\\
  \hline
  (1) Mixing up wave function and energy \\
  (2) Invoking dissipation \\
  (3) Using the energy-amplitude relation for EM waves \\
  (4) Confusion over how the electron regain energy when it reenters the wire \\
  (5) Thinking that transmitted part has less energy than whole \\
  (6) Confusion over the relationship between energy and wave function \\
  \hline
\end{tabular*}
\caption{\label{reasons}Reasons students may think that energy is lost in tunneling.  Reasons 1-3 are discussed in previous literature, and reasons 4-6 are new to the current study.}
\end{table}

However, energy loss in tunneling continued to be an issue.  After the instruction described in Section III, students asked repeatedly in lecture, problem-solving sessions, and online participation homework, ``Why is the total energy the same after tunneling?'' (These were often the same students who had given correct and clear answers to this question in earlier homework.)  In questions about tunneling that did not directly ask about energy loss, some students continued to give answers that implied that energy is indeed lost in tunneling.  Table~\ref{reasons} lists all the reasons we have seen students give for energy loss in interviews and in responses to an essay question on an exam.  Reasons (1) and (2) are the standard reasons that have been given in most previous literature, as discussed in Section III.  After describing a question we asked to elicit the idea of energy loss and the results, we will discuss examples of reasons 4-6 from interviews and observations of students in problem-solving sessions.

\begin{figure}[htbp]
  \includegraphics[width=\columnwidth]{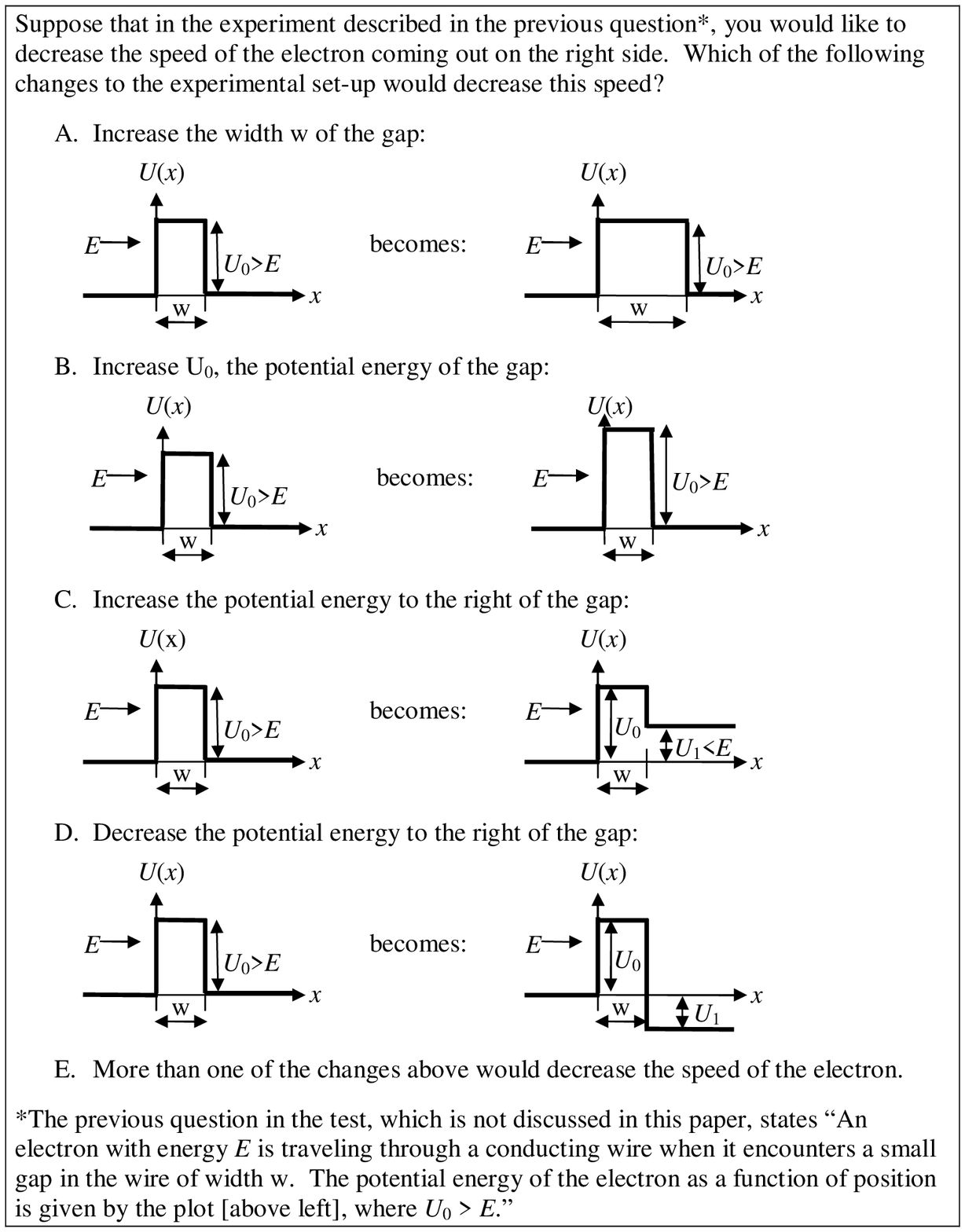}
  \caption{\label{energyloss}A tunneling question from an early version of the QMCS.  This question was developed to explore the belief that energy is lost in tunneling.  It has been removed from the QMCS because interviews suggest that responses from students in transformed courses are not necessarily indicative of whether students think energy is lost in tunneling.}
\end{figure}

Figure~\ref{energyloss} shows a question that we developed to explore the idea that energy is lost in tunneling.  In order to answer this question correctly, students must recognize that total energy is constant and determine that since potential and kinetic add up to total energy, the way to decrease the kinetic energy on the right must be to increase the potential energy, as in the correct answer, $C$.  The distracters $A$ and $B$ are very effective at eliciting the belief that energy is lost in tunneling, since students who think that energy is lost will usually think that \emph{more} energy is lost in one or both of these cases.  In interviews with students in traditional modern physics courses, we found that students' answers to this question were good indicators of whether they believed that energy is lost in tunneling; all students who did not choose the correct answer expressed the belief that energy is lost in tunneling.~\cite{McKagan2006a}

However, in later interviews with students in our transformed modern physics class, we found that even students who explicitly said that energy is not lost in tunneling sometimes chose incorrect answers, often for very subtle reasons.  (Occasionally, students even argued for answers A and B by saying that no electrons will tunnel if you make the barrier sufficiently high or wide, which will reduce the speed to zero, an argument that is technically correct.)  We have also found that this question is much more difficult than other questions eliciting the idea of energy loss, with only $37-58\%$ of students answering correctly the first time they see it.  Students' ability to answer this question also varies greatly depending on context.  As shown in Table~\ref{q10results}, when the question was asked on the QMCS, an ungraded multiple choice conceptual survey that was used as a review for the final exam, $37-41\%$ of students in the transformed course for engineering majors answered correctly (higher than the scores in the traditional course for engineering majors and comparable to the scores in the traditional course for physics majors).  However, when we gave it instead as an essay question on the final exam, asking students to ``explain your reasoning,'' $58\%$ answered correctly.  When we gave it as a multiple choice question on a final exam after asking it on the QMCS and reviewing it in class, so that students were already familiar with it, $90\%$ answered correctly.

\begin{table}
\begin{tabular}{|p{5.3cm}|r|r|r|r|r||r|}
\hline
Course           &          A &          B &    {\bf C} &          D &          E & N\\
\hline
Traditional Eng. Sp05 (QMCS) &         18 &         10 &   {\bf 24} &         15 &         33 & 68\\
Traditional Phys. Sp05 (QMCS) &         19 &         11 &   {\bf 38} &          9 &         23 & 64\\
Traditional Phys. Fa05 (QMCS) &         12 &         11 &   {\bf 38} &         15 &         24 & 54\\
Traditional Phys. Fa06 (QMCS) &         13 &         13 &   {\bf 38} &          9 &         26 & 54\\
\hline
Transformed Eng. Fa05 (QMCS) &         12 &         10 &   {\bf 37} &          5 &         37 & 162\\
Transformed Eng. Fa06 (QMCS) &         20 &          6 &   {\bf 41} &          8 &         25 & 73\\
Transformed Eng. Sp07 (QMCS) &         13 &         10 &   {\bf 40} &          7 &         31 & 120\\
\hline
Transformed Eng. Sp06 (Exam essay) &          2 &          3 &   {\bf 58} &          5 &         31 & 177\\
Transformed Eng. Sp07 (Exam multiple choice, after QMCS) &          1 &          2 &   {\bf 90} &          6 &          0 & 147\\
\hline
\end{tabular}
\caption{\label{q10results}Percentage of students who selected each answer to the question shown in Figure~\ref{energyloss} in various courses.  $N$ is the number of students.}
\end{table}

Because of the subtlety of the reasons students give for their answers and the context dependence of the scores, we no longer recommend the use of this question in multiple choice format as a diagnostic.  However, we have found that it is extremely valuable for eliciting student thinking when used in an interview setting or as an essay question on exams.

In Sp06 we asked the question in Figure~\ref{energyloss} on the final exam as an essay question.  Aside from the six students who participated in case study interviews that semester, the students had not seen the question before the exam.  Analyzing student explanations for this question sheds some light on the prevalence of the standard reasons (1-2) for the belief that energy is lost in tunneling in the transformed courses.  However, one should not attach too much importance to these numbers, as they vary considerably depending on context.  For example, a much greater percentage of students answered this question correctly when it was asked as an essay question on an exam than when it was asked as a multiple choice question on an ungraded conceptual survey.  This discrepancy is likely due to two factors: students taking the exam more seriously than the ungraded survey, and the request to explain their reasoning causing them to think more carefully about their answers and resolve their confusion.  In interviews, we saw that many students initially answered with a variety of incorrect reasoning, but in the process of attempting to explain their reasoning to the interviewer, eventually came to the correct explanation.  Thus, counting students who gave a particular incorrect reasoning in their written responses will tend to underestimate how many students have this particular difficulty.

We found that only $7\%$ of all students ($31\%$ of those who answered incorrectly) explicitly said that energy is lost in tunneling in their response to this question, although a much larger percentage gave answers that implied energy loss.  $10\%$ of all students ($43\%$ of those who answered incorrectly) related the decrease in speed to the exponential decay of the wave function, implying that they were mixing up energy and wave function (reason 1).  $16\%$ of all students ($67\%$ of those who answered incorrectly) said that it requires more energy, or is harder, to tunnel through a wider or higher barrier, implying dissipation (reason 2).  $11\%$ of all students ($48\%$ of those who answered incorrectly) argued for option $A$ and/or $B$ by pointing out that the electron would be slower in the gap in these cases than in the original case.  While it is possible that some of these students simply misread the question and thought it was asking how to slow the electron inside the gap, rather than to the right of the gap, it is clear from at least some of these responses that this is not the case.  For example, ``want to decrease its KE coming out.  We can only do this by increasing the PE in order to borrow more KE from the system'' and ``a greater PE means a decrease in KE the e- will have once it merges on right side.  Both B \& C would cause this result.'' These responses could be interpreted in terms of reason 4, which will be discussed below.

Thus, in students' written responses to an exam question, we see both of the standard reasons (1-2) for believing that energy is lost in tunneling, although only in a small minority of students.  However, in interviews, students talked about energy loss in terms of different, and much more subtle, reasons than those reported in the literature.  Aside from reason 4, these reasons do not appear in the written responses, probably because students were sufficiently unsure of them that they did not want to use them as responses on a graded exam question.

Most of the following examples are from interviews with six students who participated in case studies, where they were interviewed regularly throughout the semester.  (The examples have also been corroborated with evidence from other sources, as will be discussed below.)  Two of the interviews with each of these students (12 interviews total) focused on quantum tunneling.  In the first interview on tunneling, students were asked to go over the Tunneling Tutorial that they had already completed in class and homework.  In the second interview, they were asked a series of questions about what happens to the energy and probability of tunneling for an electron approaching a barrier when the height or width of the barrier is changed, culminating in the question in Figure~\ref{energyloss}.

In the cases where students seemed to have the common problems of mixing up energy and wave function or thinking of dissipation, they usually corrected themselves without intervention from the interviewer or instructor.  An example can be seen in an interview with a student who is struggling to answer the question in Figure~\ref{energyloss}. The student begins with a typical response in which she interprets the height of the wave function as the kinetic energy:
\begin{quote}It's either one of these two [A or B]. I'm just trying to think about it. I think it's this one right here [B], because it would--the wave function would come up here and then it would drop down a little bit but then it would keep going, and the distance between it and the potential energy would be the kinetic energy, kind of ... Well, no, this is--scratch that. We'll take it out. Because the energy has to be the same on both sides...\end{quote}  After she answered the question correctly, the interviewer asked her to explain what she was thinking before.  She said:

\begin{quote}Yeah. Um, I was thinking [pause] that a lot of times when I see these, I'm thinking of the wave function on top of it [draws wave function on top of energy graph] and I'm thinking of it dropping down a certain--dropping down, like, a certain rate depending on the difference between the energy--the electron's energy and the potential energy or the width. So I think about it that way. So I was thinking, once it--if it's coming up here and it drops down a little bit, it's gonna come up here on this side. And I'm kind of thinking, maybe, like, the amplitude of the wave function had to do with energy and so its distance from this potential was the kinetic energy, or kind of could represent the kinetic energy. But then I wasn't too sure about that, because I realized I was kind of thinking of the wave function instead of the energy, so I had to, like, re-evaluate how I was thinking about it, even though it kind of still works the same.\end{quote}

It is interesting that this student instinctively thought of drawing the wave function on top of the energy graph, although this semester, aside from in the question shown in Figure~\ref{clicker}, there were no pictures of a wave function on top of an energy graph in the lectures, textbook, or simulations.  This example illustrates that eliminating such pictures, while helpful, is not sufficient to address the problem of students mixing up energy and wave functions.

Thus, in interviews, students in the transformed course were usually able to let go of the typical ideas that lead to belief in energy loss.  However, these students often did say in interviews that energy is lost in tunneling.  There are four further reasons they gave, all of which are distinct from the reasons most often given in the literature.
\\
\\
\noindent\textbf{Reason 3: Using energy-amplitude relation for EM waves}

In the transformed courses, the relationship between amplitude and energy for electromagnetic waves was very heavily emphasized in the section on the photoelectric effect in lecture, homework, and exams.  Occasionally in interviews and observations of students in the transformed courses, but never in interviews with students in the traditional courses, students pointed out this relationship, and asked why it was not the same for matter waves, or assumed that it was the same.
\\
\\
\noindent\textbf{Reason 4: How does the electron regain energy when it reenters the wire?}

One reason that some students give for energy being lost in tunneling appears to be associated with the particular physical example we use in class, that is, an electron traveling through a wire and tunneling through an air gap.  Students can easily grasp the physical mechanism by which kinetic energy is lost when it goes from the first wire into the air gap.  Earlier in the course, in the context of the photoelectric effect, we discuss the energy required to overcome the work function of the metal.  Most students seem able to apply this concept to tunneling, recognizing that the electron loses kinetic energy when it escapes the wire into the air gap.  However, students do not understand the mechanism by which the electron regains its kinetic energy when it goes into the second wire.  Therefore, while they know that energy is conserved, they express confusion over how the electron ``gets back'' the energy that went into overcoming the work function.

For example, one student, who was sufficiently bothered by this issue that she had asked her friends about it, said in an interview:
\begin{quote}The kinetic energy starts at $E$ and then it drops down, takes energy to get up, and then it jumps back up to $E$. I talked to my friends. Why the hell, they don't understand that exactly.\end{quote}  In another interview the following week, the same student brought up the issue again: \begin{quote}Yeah, because it takes energy to get out of metal, the work function. And it takes the amount of the potential energy--the barrier, this is the barrier's, so it uses that energy up and then it has a much slower--so it's going much slower. And then once it hits the other metal, hey, it's going fast again... It's just weird, a little bit. You'd think it would slow down, but it is because the potential drops to zero again and conservation of energy, all the energy goes to kinetic. But it is a weird idea for me to think about.\end{quote}

We have found that presenting students with a gravitational analogy of a ball rolling over a hill, in which the kinetic energy is lost as the ball rolls up the hill and regained as the ball rolls down the hill, quickly resolves this confusion.  In all cases in which we have presented students with this particular difficulty with a gravitational analogy, they have been quickly satisfied.  Some researchers are reluctant to use gravitational analogies in teaching quantum tunneling, for fear that they may lead to the idea that tunneling involves a particle traveling through a physical barrier like a hill and exacerbate the difficulty of thinking that energy is lost due to dissipation.  However, we have seen no evidence of such a link.  Further, as Brookes and Etkina~\cite{Brookes2006a} point out, the gravitational analogy is already inherent in the language we use, as seen in phrases such as ``potential step'' and ``potential barrier,'' and even in the word ``tunneling.''  Even if we were careful to avoid such language, there is evidence that students have a tendency to interpret graphs too literally and think that higher on a graph means higher in space, regardless of context.~\cite{Elby2000a}  The gravitational analogy is an important aspect of the expert model of tunneling.  Therefore, we argue that it is preferable to address the strengths and limitations of the gravitational analogy directly, rather than to avoid its use.

Out of six students who were interviewed extensively on the relationship between the energy and wave function in tunneling, two exhibited this difficulty.  It was also observed in students working on homework in problem-solving sessions, and may have been a factor in some of the responses to the exam question discussed above.
\\
\\
\noindent\textbf{Reason 5: Transmitted part has less energy than whole}

\begin{figure}[htbp]
  \includegraphics[width=\columnwidth]{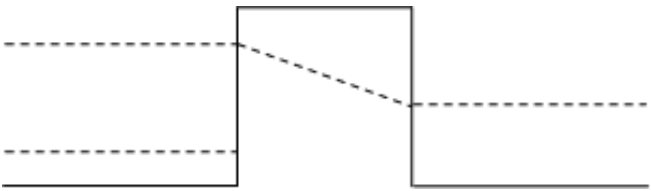}
  \caption{\label{totalenergy}A reproduction of a graph drawn by a student to represent the potential (solid line) and total (dotted lines) energy of a particle tunneling through a barrier.  This student said the particle was losing energy because only a part of it was transmitted.}
\end{figure}

Another reason students give for energy loss is that, since only part of the wave function is transmitted, only part of the energy is transmitted, with the rest being reflected.  For example, when a student working on homework during a problem-solving session was asked to draw the potential and total energy of a tunneling electron, he drew a picture like the one shown in Figure~\ref{totalenergy}.  He explained that the dotted line on the top left was the total energy of the incoming particle, which was then split into the reflected part (bottom left) and transmitted part (right).

Another student, who had explicitly stated that energy is not lost in tunneling, used a similar argument in an interview to justify her intuitive belief that the energy must be less on the right side:

\begin{quotation}
Interviewer: ...does that mean that the total energy is going down when it goes through the barrier?

Student: The total energy is constant.

I: Ah! OK, so what energy is decreasing then, if it's not the total energy?

S: The energy--- the energy--- man--- Well, OK.  What I'm saying, but what I'm saying with caution--- is--- the energy of the wave function on this side-- is decreasing.  I want to make the energy of the wave function on this side decrease.  But I'm also wary about that because--- `the energy of the wave function on this side'?  You know, the wave function is a wave function, and it has like parts to it, but it doesn't have like--- No, it does--- You can have a wave function like that… and it has a different energy here than it has here.

I: Different total energy?

S: No, total energy is of the entire wave function.  What is total energy then?  Is it this plus this plus this?  Yikes!  I need to study this for the final.
\end{quotation}

This difficulty appears to be caused by a lack of understanding of the fact that the total energy is a non-local property of the entire particle, rather than a local function of position.  This fact is not stressed in our class, nor in any class we know of, but perhaps it should be.

While this difficulty has only been observed with the two students discussed above, these two examples are from courses in different semesters, taught by different instructors using different textbooks.  Further, the problems they were working on were quite different and the two students had very different personalities.
\\
\\
\noindent\textbf{Reason 6: What is relationship between energy and wave function?}

Another reason students give for energy loss in tunneling is related to the confusion between energy and wave function (reason 1), but is more subtle.  In several interviews, we have observed students who explicitly state that the energy is not the same as the wave function, but that the two must be related somehow, so the exponential decay of the amplitude must imply a loss of energy.

For example, when asked whether the probability of tunneling would change if the width of the barrier increased, a student said some things made him think it would decrease, and other things made him think it wouldn't change.  When the the interviewer asked what made him think it wouldn't change, he said:

\begin{quote}
Well, it would be diagrams like this. [points to energy graph] One thing that the text doesn't really have--doesn't focus nearly as much as you guys do in the course, and I don't know if that's good or bad, are these diagrams. You guys use these diagrams a lot, which is great. The text doesn't so much, it sort of approaches it in a little different way. So if we are to evaluate these diagrams, put our total energy line in, evaluate how that corresponds with our potential energy, you--it sort of--[pause] maybe this forces me to think too much about energies. For example--I mean, that's more classical physics, is it not? If the particle has sufficient energy to get to the other side. Quantum's a whole other story where we're not talking about so much energies. We are, but we're also talking about probabilities, correct?  So there's sort of two ways to think about this, and maybe that's why I'm a little confused still, at this late date.
\end{quote}
Another student, when asked whether the probability of tunneling would change if the initial energy of the incoming particle decreased, said:
\begin{quote}
The amplitude shouldn't be affected by the energy other than its exposition. Yeah. I think. And then--I believe it's still gonna do the exponential decay. [draws] OK. So now--OK, so, hmm, probability of the electron tunneling through the barrier. The difference between the total energy and the potential energy of the gap is larger now, so I would say--I feel like, um, that would mean that it has less of an opportunity, less chance, less probability of it tunneling through. What am I trying to say here? When an electron has to convert a certain amount of kinetic energy to come out of a wire to potential energy, and in this case it has to convert this much [points] or this much will be potential energy, that difference there, which is more than the original case. So [pause] I don't know. Um, I'm not quite figuring out how to connect it. But the larger difference between the total energy and the potential energy of the gap I think has something to do with the probability of the electron tunneling through or not, compared to the first.
\end{quote}

This difficulty reveals why emphasizing that the wave function and energy are not the same thing is not sufficient to address the student belief that energy is lost in tunneling.  Even if students realize that energy and wave function are not literally the same thing, they struggle to make connections between these two quantities that are emphasized in the study of quantum mechanics.  One quantity, the wave function, is wholly unfamiliar to students, and the other quantity, energy, is treated in an unfamiliar way: graphed as a function of position but applied to a delocalized object.

Out of six students who were interviewed extensively on the relationship between the energy and wave function in tunneling, four exhibited some form of this difficulty.  It was also observed in students working on homework in problem-solving sessions.  Unlike the previous three difficulties, this was also observed in students in the traditional courses.  For example, one student from a traditional course, when asked how the wave function is related to the energy, replied, ``I can't remember.  I wish I did.  But I can swear, well not swear, but I can almost remember my professor saying that the energy is encoded in the wave function, somehow, I can't remember exactly now.''

\subsection{Putting Potential Energy in Context}
One conclusion of our study is that understanding the context of potential energy graphs is a difficult task for students, and a great deal of instruction is needed to address this issue.  In a previous study~\cite{McKagan2006a} we reported on interviews with students in the traditional modern physics for engineering majors course in Spring 2005.  Students in this course had no idea what the potential energy graphs mean.  In the transformed courses, we observed that students still struggled with the basic meaning of potential energy graphs, but as we refined our curriculum, their questions about these graphs became more sophisticated, illustrating a struggle to relate the graphs to physical reality in a deep and meaningful way.  The extent of questioning from students in the transformed course indicates what a difficult subject this is, and how hopeless it is to expect students to build meaningful models of these graphs if the course does not explicitly help them do so.

In Fall 2005, the first semester of the transformed course, even after focusing on the physical context of potential and explicitly addressing possible confusion arising from sloppy language in the text using ``potential'' and ``potential energy'' interchangeably, students expressed a great deal of confusion over the meaning of potential energy.  In weekly online extra credit, we asked students to submit their questions about the course material.  Here is a sample of these questions regarding potential energy:
\begin{itemize}
  \setlength{\itemsep}{1pt}
  \setlength{\parskip}{0pt}
  \setlength{\parsep}{0pt}
    \item ``I get very confused by exactly what an infinite well is. What is it, how is it infinite? do we just make it that way?''
    \item ``I have trouble understanding what the potential is when we are looking at models of an electron in a wire, free space, finite square well, infinite square well. I am sort of getting this idea of it being similar to a work function in that once the potential (V) is less than the potential energy, the electron is out of the wire. I can usually follow the math/calc that follows the examples okay, but the overall concept of this potential (V) still confuses me, and so I still don't have a firm grasp of [what] the square well models mean/represent/whatever.''
    \item ``I cant find a general description of an infinite well, i understand what it does but not what it is or where its used.''
    \item ``Voltage is used when we talk about electromagnetic forces, like the coulomb force. What I'm confused about is that we used a voltage well to show the strong force in effect. Is it accurate to show the strong force as a very deep voltage well?''
\end{itemize}

\begin{figure}[htbp]
  \fbox{\includegraphics[width=\columnwidth]{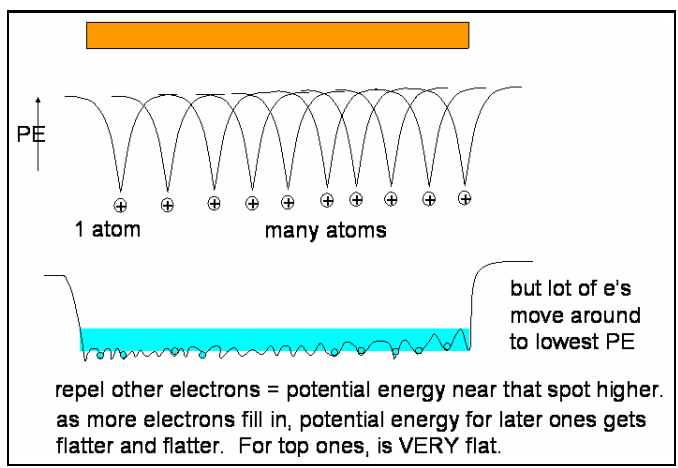}}
  \caption{\label{squarewell}An illustration of how to build up a square well by adding up the Coulomb wells of individual atoms (taken from PowerPoint slides used in lecture in Sp06, Fa06, and Sp07).}
\end{figure}

Further evidence for student confusion about potential energy can be seen in our observations of students' responses to the first question of our Tunneling Tutorial, which asked students to draw the potential energy as a function of position for an electron traveling through a long copper wire and tunneling through an air gap (see Figure~\ref{wires}b).  At this point in the course, students had worked extensively with a square well as a representation of the potential energy of an electron in a wire, but had not previously seen this example of a wire with an air gap.  We expected them to use their knowledge of the potential energy of individual wires to draw a square barrier for this new situation.  While many students did draw the correct potential energy, we observed that many students got it backwards, drawing a well instead of a barrier.  A common student explanation for the well was that the air gap was a ``hole'' and therefore should be represented by a well.  This response betrayed a lack of understanding of \textit{why} a well represents the potential energy of the wire.  In subsequent semesters, we added instruction before the Tutorial on how to build up a square well by superimposing the Coulomb potentials of all the individual atoms that make up a wire, as shown in Figure~\ref{squarewell}.  After this instruction, anecdotal observations indicated that fewer students drew a well instead of a barrier in the Tutorial.

Students still struggled to relate the potential energy graph to reality, to the extent that some viewed the graph and the electron in the wire as describing two different things.  Here is an example from an interview in which a student was trying to figure out how the width of the barrier affects the probability of tunneling:
\begin{quotation}
But I don't know if it explained as well as it needed to or if I just didn't understand as well as I needed to whether width [holds out thumb and forefinger to indicate width of space between them] meant actual real classical physics width [holds hands out to indicate width of space between them] or more theoretical width [points to potential barrier on sheet], which is like the--which might be more represented here.
\end{quotation}
Another example of a student struggling with potential energy graphs can be seen in a student who asked a question after class that revealed that he was misinterpreting the pictures in Figure~\ref{wires} as meaning that the wire was sitting on top of the potential energy.

Students also worried about the applicability, limitations, and relevance of the model of the square well for an electron in a wire.  For example, students frequently asked about collisions with the atoms in the wire, and whether these would constitute measurements of the electron and localize it.  In discussing tunneling, they struggled with the concept of infinitely long wires, and frequently discussed the reflection of the electron when it reached the end of the wire.  While working through the tunneling tutorial, one student asked why the electron would flow from one wire to the other if there was no potential difference between the two wires.  The answer to this question is that you would not have a net flow of electrons from left to right without a potential difference, but that electrons would constantly flow back and forth due to thermal energy.  This student also asked whether you could really measure a single electron flowing through a wire and why we were studying it if you couldn't.  He was satisfied only after a long explanation of how you could predict net current by adding up the effects of single electrons.  This example demonstrates that even with a physical context, a square barrier with an equal potential energy on either side (the prototypical system used in the standard presentation of tunneling) is still artificial because in reality a net current does not flow without a voltage between the two sides of the barrier.  These questions further demonstrate that physical context is important, not just for giving the material relevance, but for conceptual understanding of the material itself.

\begin{figure}[htbp]
  \includegraphics[width=\columnwidth]{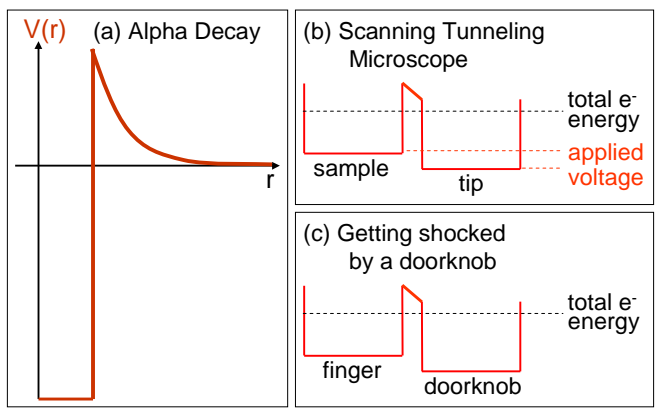}
  \caption{\label{potential}Potential energy graphs for (a) an alpha particle undergoing alpha decay, (b) an electron in a scanning tunneling microscope, and (c) an electron in your finger when you get shocked by a doorknob.  Determining how to draw each of these graphs requires many subtle approximations.}
\end{figure}

Student difficulties with potential energy can also be seen in the questions they asked during the section on the applications of tunneling, which included alpha decay, scanning tunneling microscopes, and getting shocked when you rub your foot on the carpet and approach a metal doorknob.  We asked students to figure out the potential energy graphs for each of these applications, shown in Figure~\ref{potential}, using a series of concept questions in lecture, as well as more detailed questions in homework.  Determining each of these potential energy graphs require many subtle approximations, which may not be apparent until one is faced with a barrage of student questions.  For example, to determine the potential energy graph for alpha decay, one must approximate the strong force as a flat potential throughout the nucleus, although there is no model in nuclear physics that predicts such a potential, one must recognize that the strong force dominates in the nucleus and the Coulomb force dominates outside, and one must treat the alpha particle that is going to be ejected as having an independent existence and a well-defined energy prior to decay.  Gurney and Condon~\cite{Gurney1929a} explicitly discussed all of these approximations in their 1929 paper explaining radioactivity on the basis of tunneling.  Yet most textbooks simply give such graphs without explanation.

The following questions from students illustrate that our students struggle with these approximations:
\begin{itemize}
  \setlength{\itemsep}{1pt}
  \setlength{\parskip}{0pt}
  \setlength{\parsep}{0pt}
\item ``How do the Coulomb force and the strong force relate to each other?''
\item ``How do you find the distance where the strong force takes over?''
\item ``Is the potential really square like that?''
\item ``Do alpha particles already exist in the nucleus or are they created upon radioactive decay?''
\end{itemize}
In the first two questions, students are struggling with the assumption that the strong nuclear force dominates in the nucleus, and the Coulomb force dominates outside of it.  The last two questions illustrate the simplifications required to come up with a solvable model.

Similar questions illustrated students' struggles to understand the potential energy graph for a scanning tunneling microscope:
\begin{itemize}
  \setlength{\itemsep}{1pt}
  \setlength{\parskip}{0pt}
  \setlength{\parsep}{0pt}
\item ``As the electrons tunnel through, isn't the sample potential energy going to drop?''
\item ``The quantum tunneling microscope can be used on any material even though not every material has a ``sea'' of electrons? Wouldn't losing an electron in a crucial covalent bond break the molecule apart?''
\end{itemize}
The answer to the first question is that the potential energy would drop if the sample were not hooked to a voltage supply to keep the voltage constant.  This student missed the function of the voltage supply, but the question illustrates that he was thinking carefully about the physical system.  He also recognized that the behavior of the electrons could actually change the overall potential energy, a fact which is never discussed in the standard presentation, where the potential energy function is taken as a given.  The answer to the second question is that scanning tunneling microscopes do not work on insulators, an issue that is never discussed in modern physics courses, but is the focus of a recent \emph{Nature} article~\cite{Pethica2001a}.

In spite of all these difficulties throughout the course, when we asked students to explain the physical meaning of the potential energy graph of a square barrier on a homework question towards the end of the last two semesters, nearly all gave clear and correct explanations and related the graphs to a real physical context.

Further, from interviews with 24 students in the transformed courses, there was only one case in which a student treated the potential energy graph as an external thing unrelated to the potential energy of the electron, as we saw consistently in interviews with students in a traditional course in an earlier study~\cite{McKagan2006a}.  This case was so exceptional, especially because it was a particularly good student (he received an A- in the course), that the interviewer asked him afterwards if he had done the Tunneling Tutorial.  He said he had been busy that week and skipped it, and jokingly commented, ``In conclusion, that's a good assignment, because you should listen to this guy try to explain it!''.

\subsection{Plane waves}
Plane waves cause further barriers to student understanding.  While plane waves are mathematically simple, conceptually it is quite difficult to imagine a wave that extends forever in space and time, especially when it is tunneling.  The language we use to describe tunneling is time-dependent.  For example, we say that a particle approaches a barrier from the left, and then part of it is transmitted and part of it is reflected.  This language is difficult to reconcile with a picture of a particle that is simultaneously incident, transmitted, and reflected, for all time.  The following student quote, from a homework question asking what questions students still had about tunneling after instruction, illustrates the kind of confusion created by using plane wave solutions:
\begin{quote}Say you have two finite lengths of wire very close together. I don't really see how we assume the electron is in one wire, get a solution, then use that to determine psi across the gap, and then use that to determine the probability that the electron is in the other wire. Over time don't the probabilities even out (i.e. we have no clue which wire the electron's in)?\end{quote}
This student is actually struggling with two common issues for students: confusion over the physical meaning of plane waves, and concern over what happens when the electron gets to the end of the wire.  Many students have trouble with the idea of wires extending to infinity, and talk about the electron waves reflecting off the end of the wire, interfering with themselves, and creating a big mess.  This is physically accurate, but outside of the realm of standard treatment, which assumes that wires do not have ends.

In student interviews to test the usability and effectiveness of the \emph{Quantum Tunneling and Wave Packets} simulation, we saw that students were much more comfortable with the wave packet representation than the plane wave representation.  We conducted interviews with six students, all of whom had completed either a transformed (2) or traditional (4) course in modern physics.  These students were asked to explore the simulation and think out loud.  The interviews started with free exploration, followed by questions from the interviewer about aspects of the simulation that the students had not explored on their own.  All students discovered plane wave mode on their own (the simulation starts in wave packet mode), but four out of six switched back to wave packet mode immediately and the other two only explored it for a few minutes before switching back.  Two students switched back without comment, one commented that the plane wave was ``too unrealistic,'' one commented that he didn't remember what a plane wave was and was more familiar with a wave packet, and one commented, ``That's definitely a visualization I didn't think of.''  Only one student commented that plane wave mode made sense.  Some students did eventually return to plane wave mode in order to explore specific features, but all students spent most of the free exploration time in wave packet mode, and quickly returned to wave packet mode after answering the interviewer's questions about plane wave mode.  One student, after trying it and switching back without comment, who didn't use plane wave mode again until the interviewer asked him to explore it, said he had forgotten about it.

Because the \emph{Quantum Tunneling and Wave Packets} simulation provides such a compelling visual representation, it immediately brings to the surface several troubling issues regarding plane waves that are swept under the rug in standard treatments of tunneling because textbooks focus on only a few special cases in which these issues are not apparent.

\begin{figure}[htbp]
  \fbox{\includegraphics[width=\columnwidth]{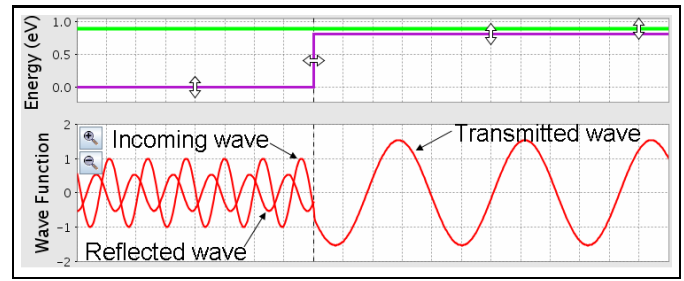}}
  \caption{\label{transmitted}A case where the amplitude of the transmitted wave is higher than the amplitude of the incident wave.}
\end{figure}

In nearly every case in quantum mechanics, with the exception of plane waves in regions of varying potential, the amplitude of the wave function gives a reasonable visual cue about relative probabilities.  In all cases other than for plane waves, we teach that the probability density is, by definition, given by:
\begin{equation}\label{probother}
    P(x) = |\psi(x)|^2
\end{equation}
Thus, it seems reasonable that relative amplitudes should be a good indication of relative probabilities.  However, this is not necessarily the case for plane waves, and the simulation reveals that there are even cases in which the amplitude of the transmitted wave is larger than the amplitude of the incident wave (see Figure~\ref{transmitted}).  This is such a surprising result that many experts, when they first see such a case, think there is a bug in the simulation.

Students often cue off the amplitude of the plane wave as a measure of probability and draw incorrect conclusions.  In observations of students attempting to calculate reflection and transmission coefficients during problem-solving sessions, we noticed that many students initially assumed that they were given by:
\begin{equation}\label{wrongR}
R=|B|^2/|A|^2\end{equation}
and
\begin{equation}\label{wrongT}
T=|C|^2/|A|^2\end{equation}
where $A$, $B$, and $C$ are the amplitudes of the incident, reflected, and transmitted waves, respectively.  These equations happen to be correct for plane waves tunneling through a square barrier with the same potential on both sides, since the particle speeds happen to cancel, but Equation~\ref{wrongT} is wrong for a step potential or for any other situation in which the potential is different for the incident and transmitted waves.

Thus, both faculty and students tend to assume that the amplitude alone is an accurate indicator of probability, as in Equation~\ref{probother}, and make mistakes as a result.  Yet most textbooks quickly gloss over this issue.  Most quantum mechanics textbooks simply state that the reflection and transmission coefficients for plane waves are determined by the probability current, without explaining why it is necessary to introduce this concept here and not elsewhere.  In many modern physics textbooks, this issue is not discussed at all, and the equation for the transmission coefficient is simply given, either in terms of particle velocities ($v$) or wave numbers ($k$),
\begin{equation}\label{transmission}
    T = \frac{v_t|C|^2}{v_i|A|^2} = \frac{k_t|C|^2}{k_i|A|^2}
\end{equation}
(the subscripts $i$ and $t$ denote the incident and transmitted waves, respectively), with no explanation of where the factors of $k$ or $v$ come from.  Some textbooks simply give Equation~\ref{wrongT} for the transmission coefficient, with no mention that this applies only for the special case in which $v_t=v_i$.~\cite{Griffiths2005a,Taylor2004a,Krane1996a,Serway2005a,Ohanian1995a,Knight2004a}  While we understand that the authors of these textbooks are attempting to avoid excessive mathematics that would obscure the basic concept of tunneling, presenting the transmission coefficient only for this special case leads students to draw many incorrect conclusions when attempting to extend their knowledge to other contexts.  One textbook even writes down Equation \ref{wrongR} as the obvious expression for R, and then ``derives'' the correct expression for T (Equation~\ref{transmission}) by stating that it follows from the ``convention'' that $R+T=1$.~\cite{Eisberg1985a}

We know of only two textbooks that give further justification by deriving the probability for a wave packet in the limit that the width goes to infinity.~\cite{Cohen-Tannoudji1977a,Shankar1994a}  However, even in these books, it is not intuitively clear \emph{why} an infinitely wide wave packet should lead to a probability proportional to the particle speed.  We recommend an alternative treatment suggested by Lande et al.~\cite{Lande2008a}, in which reflection and transmission coefficients are derived from wave packets, demonstrating that the factor of the $v$ results from the fact that the widths of the reflected and transmitted wave packets are a function of the speed at which they move in their respective media.  This derivation is more intuitive than the derivation from probability current, both because it relates more easily to the typical definition of probability as it relates to the amplitude of the wave function, and because wave packets are more physical than plane waves.

A second problematic issue that is often swept under the rug is the issue of wave speed vs. particle speed.  Because the treatment of waves is being pushed out of the physics curriculum at many institutions, many students do not know the difference between phase velocity ($v_\phi=\omega/k$) and group velocity ($v_g=d\omega/dk$).  For a Schr{\"o}dinger wave function, the phase and group velocity are given by:
\begin{eqnarray}
  v_\phi &=& \frac{\hbar k}{2m} + \frac{V}{\hbar k} \\
  v_g &=& \frac{\hbar k}{m}
\end{eqnarray}
While the velocity of a particle corresponds to the group velocity of its wave function, the only velocity apparent in the visual representation of a plane wave is the phase velocity.  The distinction causes confusion when the potential energy changes.  Students can see in the simulation that if they increase the potential energy, the ``wave speed'' increases, which seems to contradict their intuition that increasing the potential energy should decrease the kinetic energy, and therefore the speed (since $KE = E - V$).  In fact, increasing the potential energy \emph{increases} the phase velocity, or wave speed, but \emph{decreases} the group velocity, or particle speed.  The only way we know to gain any physical intuition for the group velocity of a plane wave is again to imagine it as an infinitely wide wave packet, in which case the group velocity is the speed at which that wave packet travels.

The distinction between wave speed and particle speed also causes problems in trying to explain why the probability is not proportional only to the square of the amplitude of the wave function.  As discussed above, the transmitted amplitude can be larger than the incident amplitude if the transmitted particle speed is smaller.  However, in all such cases, the wave speed is actually larger, so it appears that the transmitted wave has larger amplitude and is moving faster, obscuring the correct explanation, that it has a smaller particle speed to compensate for the larger amplitude.

We point out these issues so that instructors will be aware of the complexities inherent in discussing plane waves and consider the advantages of focusing on more realistic wave packets.  We do not have solutions for how to address the difficulties with plane waves (aside from avoiding plane waves and focusing on wave packets), and we hope that other researchers will pursue these questions further.

\subsection{Representations of complex wave functions}

Students often have difficulty understanding the meaning of complex wave functions.  This can perhaps best be illustrated by the observation that students frequently ask, ``What is the physical meaning of the imaginary part of the wave function?'' but never ask about the physical meaning of the real part, even though both have the same physical significance.

Ambrose~\cite{Ambrose1999a} found that some students believe that the wave function is only ``real'' in classically allowed regions, so that the real part is zero inside the barrier.  We saw this problem in one interview.

All of the textbooks used in courses in this study regularly plotted only the real part of the wave function, but referred to it as ``the wave function,'' as in Figure~\ref{planewave}.  In the QMCS and in the transformed courses, we always labeled such pictures explicitly as the real part.  However, we found in interviews that even students in the transformed courses who had seen explicit discussion of both the real and imaginary parts were often confused by requests to draw ``the real part of the wave function.''  When asked to draw the real part of the wave function on the exam question discussed at the beginning of Section VI, 6 students ($3\%$) said that the wave function is only real inside the barrier and set it to zero everywhere else.

\begin{figure}[htbp]
  \fbox{\includegraphics[width=\columnwidth]{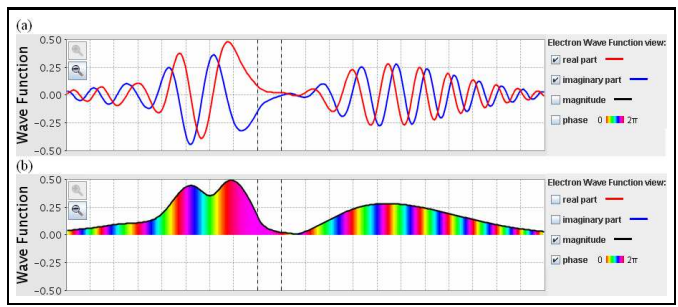}}
  \caption{\label{representation}(a) A representation showing the real and imaginary parts of a wave function and (b) a representation showing the magnitude and phase of a wave function.  In interviews we see that students can make sense of representation (a) but not representation (b).}
\end{figure}

To address these problems, we designed the \emph{Quantum Tunneling and Wave Packets} simulation (as well as two other PhET simulations on quantum wave functions, \emph{Quantum Bound States} and \emph{Quantum Wave Interference}) to include both the real and imaginary parts on an equal footing (see Figure~\ref{representation}a), and to include time dependence so that students could see how the wave function alternates in time between the real and the imaginary parts.  For completeness, we also included the ``phase color'' representation used exclusively in most non-PhET simulations of wave functions, in which a curve representing the magnitude of the wave function is filled in with colors representing the phase (Fig.~\ref{representation}b).

In interviews with five students on \emph{Quantum Wave Interference}~\cite{QWIsim}, one student commented that he did not understand real and imaginary numbers, and one student wondered why the imaginary part didn't look different from the real part until he paused the simulation and could see that they were out of phase.  Aside from these two, whose confusion stemmed more from their expectations than from the simulation, the students interviewed did not express any confusion over the real and imaginary representations of the wave function in interviews on \emph{Quantum Tunneling and Wave Packets} and \emph{Quantum Wave Interference}.  Several students also learned important concepts by playing with the real and imaginary views.  For example, students figured out from the simulation that the real and imaginary parts were 90 degrees out of phase, and that the real and imaginary parts add up to a constant probability density in an energy eigenstate even though each individual component changes in time.

On the other hand, the ``phase color'' representation caused significant problems for most students.  In interviews on \emph{Quantum Wave Interference}, three out of five students interviewed explored this view.  None of the three made any comments on it on their own, aside from one student who said it hurt his eyes, so the interviewer asked them what it was showing.  One student said it was ``some sort of frequency type of thing'' and speculated that teal would constructively interfere with teal and destructively interfere with the opposite of teal.  Another stared at the screen in confusion for over a minute, and then described it as ``some sort of representation of both the real part and the imaginary part'' showing that ``pink is areas of high real part and low imaginary part or something?''  Another student was unable to give any explanation.  When the same three students were interviewed later on \emph{Quantum Tunneling and Wave Packets}, the two who had given explanations in earlier interviews did not comment on phase view again.  The student who had been unable to give any explanation remembered that this view had been used in his quantum course, but still could not explain what it meant.  Of three additional students who were interviewed on \emph{Quantum Tunneling and Wave Packets} but not \emph{Quantum Wave Interference}, two expressed frustration over the phase view and were unable to explain it, and the third, when asked to explain it, said only that it showed ``something about wavelength.''  When given a choice, none of the students spent much time in phase mode, returning quickly to real or magnitude mode after answering the interviewer's questions.

``Phase color'' is still an option in the simulations for instructors who would like to explicitly teach the use of this representation or use activities developed for other simulations.  However, based on our interviews, we do not recommend the use of the ``phase color'' representation with students.

\subsection{``Hard Questions''}
One striking result of our transformed instruction was the number of student questions probing the relationship of the course material to reality, many of which were sufficiently difficult that most expert physicists could not easily answer them.  Many examples of these questions have already been discussed in Section VIB.  Below are some further examples:
\begin{itemize}
  \setlength{\itemsep}{1pt}
  \setlength{\parskip}{0pt}
  \setlength{\parsep}{0pt}
    \item What [happens if the electron is spread out] in the wire, and you cut the wire in half?
    \item How come we don't count the position in the wire?  How come we only count the energy?
    \item Wouldn't there be a charge difference in the wire if it were more likely to be found in the center?
    \item If everything's got to be measured for it to be localized, how come everything's already localized?  I'm not going around measuring things.
\end{itemize}
We hypothesize that these questions are a result of the combination of interactive engagement techniques with a focus on real world applications.  Our students are constantly engaged in a struggle to relate the material to reality.  We regard the quantity of such questions as a sign that this struggle is very difficult.  We question whether there is much learning in courses where students are not asking such questions.

\section{Lessons for improving student learning of quantum tunneling}

Our research demonstrates that a focus on addressing common student difficulties is helpful, but not sufficient, for improving student learning of quantum tunneling.  By addressing these difficulties and focusing on relating the material to reality, we have uncovered deeper problems in students' ability to use the basic models of quantum mechanics, such as wave functions as descriptions of physical objects, potential energy graphs as descriptions of the interactions of those objects with their environments, and total energy as a delocalized property of an entire wave function that is a function of position.  We have found that real world examples are useful not just to help students see the connection to their lives, but also to help them make sense of the models they are using.

Effective curriculum on quantum tunneling must explicitly help students learn to build these models.  Two practices that we have found useful are focusing on how to relate potential energy graphs to physical systems and starting with wave packets rather than plane waves.

There are several further practices that, although we have not tested them on a large scale, our research suggests would be valuable.  These include:
\begin{enumerate}
  \setlength{\itemsep}{1pt}
  \setlength{\parskip}{0pt}
  \setlength{\parsep}{0pt}
    \item Tutorials to lead students through the process of drawing potential energy graphs for various physical situations.~\footnote{The Activity-Based Tutorials~\cite{Wittmann2005b}, one of which we adapted for use in our curriculum, also include several exercises for building up the idea of potential energy graphs through lab activities with carts on magnetic tracks.  We did not use these activities due to lack of time and lack of a lab section in our course, but our research indicates that such an approach could be useful, if connections are made between these classical examples and examples of systems where quantum mechanics applies.}
    \item Explicit discussion of the strengths and weaknesses of gravitational analogies.
    \item Explicit discussion of the reasons for the focus in quantum mechanics on an energy representation rather than the force representation used in introductory physics.
    \item Explicit discussion of why total energy is quantized (for bound particles), but potential energy is not.
\end{enumerate}

\section{Acknowledgments}
We thank Chris Malley, the software engineer for the \emph{Quantum Tunneling and Wave Packets} simulation, as well as Mike Dubson and Sam Reid for help with the physics and numerical methods involved in the simulation.  We thank Michael Wittmann for convincing us to write this paper, and Travis Norsen for many useful discussions and feedback.  We also thank the PhET team and the Physics Education Research Group at the University of Colorado.  This work was supported by the NSF, The Kavli Institute, The Hewlett Foundation, and the University of Colorado.

\bibliography{../bibliographies/PER}
\bibliographystyle{apsrev}
\end{document}